\PassOptionsToPackage{final}{graphicx}
\documentclass[sigconf,natbib=false,nonacm]{acmart}

\usepackage[backend=biber,bibencoding=utf8,style=ACM-Reference-Format,natbib,mincitenames=1,maxcitenames=2,maxbibnames=10,autocite=inline,sortcites,labelnumber]{biblatex}
\DeclareFieldFormat*{citetitle}{\emph{#1}}

\hypersetup{draft=false}

\usepackage[obeyDraft,colorinlistoftodos,prependcaption]{todonotes}

\usepackage[group-separator={,},group-minimum-digits={3},detect-all]{siunitx}
\DeclareSIUnit[number-unit-product = ]\percent{\char`\%}

\usepackage[export]{adjustbox}
\usepackage{nicefrac}
\usepackage{textcomp}
\usepackage{soul}
\usepackage{mathrsfs}
\usepackage{amsmath}
\usepackage{amsfonts}
\usepackage{mathtools}
\usepackage[capitalize]{cleveref}

\crefname{listing}{listing}{listings}
\Crefname{listing}{Listing}{Listings}
\crefname{sublisting}{listing}{listings}
\Crefname{sublisting}{Listing}{Listings}

\usepackage{comment}
\usepackage{ellipsis}

\usepackage{algorithm}
\usepackage{algorithmic}

\crefname{ALC@unique}{line}{lines}
\Crefname{ALC@unique}{Line}{Lines}

\crefname{paragraph}{section}{sections}
\Crefname{paragraph}{Section}{Sections}

\graphicspath{{./graphics/}} %

\usepackage{listofitems}
\usepackage{numname}
\usepackage[super]{nth}

\usepackage[shortlabels,inline]{enumitem}

\usepackage[font=small,skip=0.75em]{caption}
\setlist{noitemsep,nosep}

\makeatletter
\newcommand\notsotiny{\@setfontsize\notsotiny\@vipt\@viipt}
\makeatother

\usepackage[
    draft=false,
    newfloat,
    frozencache,
    cachedir=.
]{minted}
\newminted{diff}{
    numbersep=3pt,
    style=trac,
}
\newminted{python}{}
\newmintinline{python}{}
\definecolor{bg}{rgb}{0.95,0.95,0.95}
\newminted{pycon}{bgcolor=bg,linenos=false,tabsize=2}
\newmintinline{pycon}{}
\newmintinline{text}{
}
\setmintedinline{
    style=bw,
    fontsize=\small}
\setminted{%
    numbersep=4pt,
    style=trac,
    fontsize=\scriptsize,
    autogobble,
    breaklines,
    resetmargins,
    breakindentnchars=2,
    linenos,
    tabsize=2,
    breakafter=.,
    breakautoindent,
    breakaftersymbolpre=,
    escapeinside=||}

\newcommand{\pyi}[1]{\pythoninline{#1}}
\newcommand{\pyc}[1]{\pyconinline{#1}}

\usepackage{booktabs}
\usepackage{tabulary}
\usepackage{tabularx}
\usepackage{threeparttable}
\usepackage{multirow}

\newcommand{\NumProjects}{\num{250}}

\newcommand{\MLOC}{\num{19.7}}
\newcommand{\NumCommits}{\num{199140}}
\newcommand{\NumIssues}{\num{237232}}
\newcommand{\NumYears}{\num{460.21}}
\newcommand{\AvgYears}{\num{1.86}}

\newcommand{\NumCommitsManuallyExamined}{\num{470}}
\newcommand{\NumIssuesManuallyExamined}{\num{446}}
\newcommand{\NumManuallyExamined}{\num{916}}

\newcommand{\NumBugs}{157}
\newcommand{\NumDifficulties}{123}
\newcommand{\NumChallenges}{280}
\newcommand{\NumRefactTang}{\num{285}}
\newcommand{\NumBugsZhang}{\num{175}}
\newcommand{\NumBugsKhatchadourian}{\num{61}}

\newcommand{\UnknownCategoriesPercentage}{\SI{3.57}{\percent}}
\newcommand{\PerformancePercentage}{\SI{39.64}{\percent}}
\newcommand{\PerformanceFraction}{\nicefrac{111}{280}}
\newcommand{\RemovedTFFPercentage}{\SI{7.21}{\percent}}
\newcommand{\RemovedTFFFraction}{$\nicefrac{8}{111}$}
\newcommand{\PerfFixedByAddingTFFPercentage}{\SI{54.95}{\percent}}
\newcommand{\PerfFixedByAddingTFFFraction}{$\nicefrac{61}{111}$}
\newcommand{\PerfFixedOtherPercentage}{\SI{45.05}{\percent}}

\newcommand{\PerfFixedByArgsPercentage}{\SI{25.23}{\percent}}
\newcommand{\InputShapePercentageOfPerformanceProblems}{\SI{18.92}{\percent}}
\newcommand{\APIMisusePercentage}{\SI{18.93}{\percent}}
\newcommand{\APIConfusionPercentage}{\SI{37.74}{\percent}}
\newcommand{\APIConfusionFraction}{$\nicefrac{20}{53}$}
\newcommand{\APIMisuseFixedByRemovalPercentage}{\SI{28.30}{\percent}}
\newcommand{\APIMisuseFixedByRemovalFraction}{$\nicefrac{15}{53}$}
\newcommand{\IncompatabilityPercentage}{\SI{17.50}{\percent}}
\newcommand{\IncompatabilityFraction}{$\nicefrac{49}{280}$}
\newcommand{\RuntimeErrorUnexpectedBehabiorIncompatabilitySymptomPercentage}{\SI{81.63}{\percent}}
\newcommand{\TensorFlowBugsPercentage}{\SI{7.86}{\percent}}
\newcommand{\TensorFlowBugsAsIssuesPercentage}{\SI{81.82}{\percent}}

\newcommand{\PercentagOfTensorFlowBugsLeadingToDeadlock}{\SI{9.09}{\percent}}

\newcommand{\TopLevelCategories}{\num{12}}
\newcommand{\OverallCategories}{\num{24}}

\newcommand{\TFFReleaseDate}{September 30, 2019}

\newenvironment{commit}[1]{
    \setsepchar{, }
    \readlist\args{#1}
    \begin{listing}
}{
	\vspace{-0.25em}
	\caption{\small Commit \args[2] in \args[1]: \args[9]}
	\label{lst:\args[10]}
	\vspace{-1em}
    \end{listing}
}

\usepackage{mdframed}
\newcounter{finding}
\newmdenv[%
    linecolor=black,
    outerlinewidth=0pt,
    skipabove=0pt,
    skipbelow=0pt,
    settings={\global\refstepcounter{finding}},
]{myfinding}

\newcommand{\finding}[1]{
    \begin{myfinding}
	\vspace{-0.25em}
	\textbf{\textit{Finding~\arabic{finding}}}: #1
	\vspace{-0.25em}
    \end{myfinding}
}

\crefname{rq}{}{}
\Crefname{rq}{}{}

\crefname{finding}{finding}{findings}
\Crefname{finding}{Finding}{Findings}

\newlist{bestpractices}{enumerate}{1}
\setlist[bestpractices]{label=(\arabic*),ref=\arabic*} %
\crefname{bestpracticesi}{best practice}{best practices}
\Crefname{bestpracticesi}{Best practice}{Best practices}

\newlist{antipatterns}{enumerate}{1}
\setlist[antipatterns]{label=(\arabic*),ref=\arabic*} %
\crefname{antipatternsi}{anti-pattern}{anti-patterns}
\Crefname{antipatternsi}{Anti-pattern}{Anti-patterns}

\newlist{recommendations}{enumerate}{1}
\setlist[recommendations]{label=(\arabic*),ref=\arabic*} %
\crefname{recommendationsi}{recommendation}{recommendations}
\Crefname{recommendationsi}{Recommendation}{Recommendations}

\newcommand{\GH}{\href{http://github.com}{GitHub}}

\newcommand{\TF}{\href{http://github.com/tensorflow/tensorflow}{\textinline{TensorFlow}}}
\newcommand{\galference}{\href{https://github.com/modichirag/galference}{\textinline{galference}}}

\newcommand{\addons}{\href{https://github.com/tensorflow/addons}{\textinline{tensorflow/addons}}}
\newcommand{\neuro}{\href{https://github.com/MLH-Fellowship/neuro-art}{\textinline{neuro-art}}}
\newcommand{\DDPG}{\href{https://github.com/samuelmat19/DDPG-tf2}{\textinline{DDPG-tf2}}}
\newcommand{\GPflow}{\href{https://github.com/GPflow/GPflow}{\textinline{GPflow}}}

\makeatletter
\def\TFF{\@ifstar\@TFF\@@TFF}
\def\@TFF{\pythoninline{tf.function}}
\def\@@TFF{\pythoninline{@tf.function}}
\makeatother

\setcopyright{acmcopyright}
\copyrightyear{2022}
\acmYear{2022}
\acmDOI{10.1145/3524842.3528455}
\acmISBN{978-1-4503-9303-4/22/05} %
\acmPrice{15.00}

\acmConference[MSR '22]{19th International Conference on Mining Software Repositories}{May 23--24, 2022}{Pittsburgh, PA, USA}
\acmBooktitle{19th International Conference on Mining Software Repositories (MSR '22), May 23--24, 2022, Pittsburgh, PA, USA}

\begin{document}

\title[Challenges in
Migrating Imperative Deep Learning Programs to Graph Execution]{Challenges in
Migrating Imperative Deep Learning Programs to Graph Execution: An Empirical Study}

\author{Tatiana Castro V\'{e}lez}
\orcid{0000-0002-6122-269X} %
\affiliation{%
  \institution{City University of New York (CUNY) Graduate Center}
  \city{New York}
  \state{NY}
  \country{USA}
}
\email{tcastrovelez@gradcenter.cuny.edu}

\author{Raffi Khatchadourian}
\orcid{0000-0002-7930-0182} %
\affiliation{%
  \institution{City University of New York (CUNY) Hunter College}
  \city{New York}
  \state{NY}
  \country{USA}
}
\email{raffi.khatchadourian@hunter.cuny.edu}

\author{Mehdi Bagherzadeh}
\orcid{0000-0003-1549-881X} %
\affiliation{%
  \institution{Oakland University}
  \city{Rochester}
  \state{MI}
  \country{USA}
}
\email{mbagherzadeh@oakland.edu}

\author{Anita Raja}
\orcid{0000-0002-0735-7358} %
\affiliation{%
  \institution{City University of New York (CUNY) Hunter College}
  \city{New York}
  \state{NY}
  \country{USA}
}
\email{anita.raja@hunter.cuny.edu}

\begin{abstract}

    Efficiency is essential to support responsiveness w.r.t.~ever-growing datasets, especially for
    Deep Learning (DL)
    systems. DL frameworks
    have traditionally embraced \emph{deferred} execution-style DL code that supports symbolic, graph-based Deep Neural Network (DNN) computation. While scalable, such development
    tends to produce DL code that is error-prone, non-intuitive, and difficult to debug. Consequently, more natural, less error-prone imperative DL frameworks encouraging \emph{eager} execution have emerged
    at the expense of run-time performance. While hybrid approaches aim for
    the ``best of both worlds,''
    the challenges
    in applying them in the real world are largely unknown. We
    conduct a data-driven analysis of challenges---and resultant bugs---involved in writing reliable yet performant imperative DL code by
    studying
    \NumProjects\ open-source projects, consisting of \MLOC\ MLOC, along with \NumCommitsManuallyExamined\ and \NumIssuesManuallyExamined\ manually examined code patches and bug reports, respectively. The results indicate that
    hybridization:
    \begin{enumerate*}[(i)]
	\item is prone to API misuse,
	\item can result in performance \emph{degradation}---the opposite of its intention, and
	\item has limited application due to execution mode incompatibility.
    \end{enumerate*}
    We put forth several recommendations, best practices, and anti-patterns for effectively hybridizing imperative DL code, potentially benefiting DL practitioners, API designers, tool developers, and educators.

\end{abstract}

\ccsdesc[500]{General and reference~Empirical studies}
\ccsdesc[500]{Computing methodologies~Machine learning}
\ccsdesc[300]{Software and its engineering~Language features}
\ccsdesc[300]{Software and its engineering~Software evolution}

\keywords{empirical studies, deep learning, imperative programs, hybrid programming paradigms, graph-based execution, software evolution}

\listoftodos{}

\maketitle

\section{Introduction}\label{sec:intro}

Machine Learning (ML), including Deep Learning (DL), systems are pervasive in
society. Central to
such
systems are dynamic
models, whose behavior
is ultimately defined by input data. However, as datasets grow, efficiency becomes essential to support
responsiveness~\cite{Zhou2020}. For industrial applications, DL frameworks---pillars of DL systems~\cite{Liu2020,Islam2019,Zhang2018,Islam2020a}---must quickly execute complex computations on large datasets while
supporting easy-to-use programming paradigms~\cite{Jeong2019}.
For efficiency, DL frameworks have traditionally embraced a \emph{deferred} execution-style
that supports symbolic, graph-based Deep Neural Network (DNN) computation~\cite{Google2021,Chen2015}. While scalable,
development is error-prone, cumbersome, and produces programs that are difficult to debug~\cite{Zhang2018,Islam2019,Islam2019a,Zhang2019}.
Furthermore, because graph computation executes statements in a non-imperative order, traditional Software Engineering (SE) tools cannot help troubleshoot bugs~\cite{Arpteg2018}. Contrarily, more natural, less error-prone, and easier-to-debug \emph{imperative} DL frameworks~\cite{Agrawal2019,Paszke2019,Chollet2020} encouraging \emph{eager} execution have emerged. Though ubiquitous, eagerly-executed imperative DL programs are less efficient and scalable as their deferred-execution counterparts~\cite{Chen2015,Paszke2019,Moldovan2019,Facebook2019,Jeong2019,Google2022}.
Executing (imperative) DL programs eagerly ``makes tensor [matrix-like data structures central to DL] evaluation trivial but at the cost of lower performance''~\cite{Marcinkiewicz2021}.\footnote{Performance is this paper refers to \emph{run-time} performance (speed), not model accuracy.} Thus,
hybrid approaches~\cite{Moldovan2019,Facebook2019,Apache2021b}---integrated into mainstream DL frameworks---execute
imperative DL programs as static graphs at run-time.
For example, in \citetitle{Abadi2016}~\cite{Abadi2016}---a popular~\cite{Zhang2018,He2019}
DL framework---\citetitle{Moldovan2019}~\cite{Moldovan2019} can
potentially enhance
performance by decorating (annotating)---with optional yet influential decorator arguments---appropriate Python function(s) %
with \TFF. Decorating functions with such hybridization Application Programming Interfaces (APIs) %
can increase
imperative DL code performance without explicit modification.

Though promising, hybrid approaches
necessitate
non-trivial
specialized metadata~\cite{Jeong2019} and exhibit
limitations
and known issues~\cite{Google2021b}
with
native program constructs.
Subtle considerations are required to make
code amenable to safe, accurate, and efficient graph execution---avoiding performance bottlenecks and semantically inequivalent results. Therefore,
developers are burdened with making their
code compatible with the underlying execution model conversion, as well \emph{manually} specifying which
functions should be converted.
While alternatives~\cite{Jeong2019}
exist, they impose custom
Python interpreters, which may be impractical for
industry, and support only specific Python constructs. Thus, there is a knowledge gap in how hybridization
is
used in real-world
DL applications,
leading to the challenges
in successfully applying
it underexplored. Without such insight, DL systems may be inefficient, fallible, and difficult to maintain. Moreover,
advances in DL
are likely to be futile if
they
cannot be effectively used.

To fill this gap, we conduct an empirical study on common development challenges in migrating imperative
DL code to graph execution using hybridization
in open-source DL systems. Particularly, we aim to answer the following research questions:
\begin{enumerate*}[(RQ1)]
    \item what bug patterns and corresponding challenges are involved in writing \emph{reliable} yet \emph{performant} imperative DL code,\label[rq]{rq:bugs} and %
    \item which best practices and anti-patterns can be extracted from \cref{rq:bugs}\label[rq]{rq:ex}? %
\end{enumerate*}
Such knowledge can help drive new automated migration techniques, IDE code completion, and automated (data science-specific~\cite{Dilhara2021,Dilhara2022,Atwi2021}) refactoring mining approaches~\cite{Tsantalis2020}.
The results:
 \begin{enumerate*}[(i)]
     \item advance knowledge of this emerging yet pervasive hybrid paradigm,
     \item provide feedback to language and API designers for future API versions,
     \item help tool designers comprehend difficulties with writing performant imperative DL code,
     \item include preliminary \emph{recommendations}, \emph{best practices}, and \emph{anti-patterns} for practitioners in using hybridization
	 effectively, and
     \item assist educators in teaching hybridization APIs.\label{itm:edu}
\end{enumerate*}

Our study involves analyzing
occurrences of \TFF*\ in \NumProjects\ projects, consisting of \MLOC\ MLOC, along with \NumCommitsManuallyExamined\ and \NumIssuesManuallyExamined\ manually examined code patches (Git commits) and bug reports (\GH\ issues), respectively.
Challenges---along with their causes, symptoms, and fix patterns---are
taxonomized using manual processes aided by automated software repository mining. Due to its popularity and extensive analysis by previous work~\cite{Zhang2018,Islam2020a,Islam2019,Liu2020,Nikanjam2021,Chen2021,Zhang2019,Humbatova2020},
we focus on hybridization in \citetitle{Abadi2016}.
Our study indicates that:
\begin{enumerate*}[(i)]

    \item \TFF*\ is widely used,

    \item misusing \TFF*\ was a major theme in migrating imperative DL programs to graph execution,

    \item subtle bugs in using \TFF*\ can
	result in performance \emph{degradation}---the opposite of its intention, and

    \item
	\TFF*\ is commonly incompatible in a given context---limiting its application.

\end{enumerate*}

Our contributions can be summarized as follows:

\begin{description}

    \item[Hybridization bug hierarchical taxonomy] From \NumCommitsManuallyExamined\ and \NumIssuesManuallyExamined\ patches and bug reports, respectively, of \NumProjects\ projects manually examined, we build a rich hierarchical taxonomy of common hybridization challenges.

    \item[Recommendations, best practices, \& anti-patterns] We propose preliminary recommendations, best practices, and anti-patterns for effectively hybridizing imperative DL code from our statistical results, as well as an in-depth analysis. %

\end{description}

\noindent Complete results of our study are available in our dataset~\cite{Velez2022}.

\section{Motivating Examples \&
Background}\label{sec:motive}

\begin{listing}
    \begin{minipage}[t]{0.5\linewidth}
	\begin{pythoncode}
		# Build a graph.
		a = tf.constant(5.0)|\label{lne:graph_start}|
		b = tf.constant(6.0)
		c = a * b|\label{lne:add}\label{lne:graph_end}|
	\end{pythoncode}
    \end{minipage}\hfill
    \begin{minipage}[t]{0.5\linewidth}
	\begin{pythoncode*}{firstnumber=last}
		# Launch graph in a session.
		sess = tf.Session()|\label{lne:session_create}|
		# Evaluate the tensor `c`.
		print(sess.run(c)) # prints 30.0|\label{lne:session_run}|
	\end{pythoncode*}
    \end{minipage}
    \caption{\citetitle{Abadi2016} deferred execution-style code~\cite{Google2021r}.}\label{lst:def}
\end{listing}

Popular DL frameworks have historically embraced \emph{deferred} execution-style (low-level) APIs, making DNNs straight-forward to execute as symbolic graphs
that enable various run-time optimizations. For example, during graph building (lines~\ref{lne:graph_start}--\ref{lne:graph_end} of \cref{lst:def}),
line~\ref{lne:add} does not execute until the \pythoninline{Session} created on line~\ref{lne:session_create} is run on line~\ref{lne:session_run}. While efficient, legacy code using
such APIs are cumbersome, error-prone, and difficult to debug and maintain~\cite{Zhang2018,Islam2019,Islam2019a,Zhang2019}.
Such APIs also do not natively support common imperative program constructs, e.g., iteration~\cite{Apache2018}. Contrarily, \emph{eager} execution-style DL APIs~\cite{Agrawal2019,Paszke2019} facilitating higher-level, imperative, and Object-Oriented (OO)~\cite{Chollet2020} (Python) programs that are easier-to-debug, less error-prone, and more extensible have emerged. For instance, with eager execution, line~\ref{lne:add} of \cref{lst:def} would execute and immediately evaluate tensor \pythoninline{c}, foregoing the need of a session. In many DL frameworks, eager execution is now the default.

\begin{listing}
    \begin{minipage}[t]{0.65\linewidth}
	\begin{pythoncode}
		class SequentialModel(tf.keras.Model):|\label{lne:subclass}|
			def __init__(self, **kwargs):
				super(SequentialModel, self).__init__(...)
				self.flatten = layers.Flatten(
					input_shape=(28, 28))
				num_layers = 100 # Add many small layers.
				self.layers = [layers.Dense(64, activation =
					"relu") for n in range(num_layers)]
				self.dropout = tf.keras.layers.Dropout(0.2)
				self.dense_2 = tf.keras.layers.Dense(10)
	\end{pythoncode}
    \end{minipage}\hfill
    \begin{minipage}[t]{0.35\linewidth}
	\begin{pythoncode*}{firstnumber=last}
		@tf.function(...) # Executes|\label{lne:tfunc}|
			# the model as a graph
			# (with optional args).
		def __call__(self, x):
			x = self.flatten(x)
			for layer in self.layers:
				x = layer(x)
			x = self.dropout(x)
			x = self.dense_2(x)
			return x
	\end{pythoncode*}
    \end{minipage}
    \caption{\citetitle{Abadi2016} imperative (OO) DL model code~\cite{
	Google2022}.}\label{lst:model}
\end{listing}

Despite the benefits, executing (imperative) DL programs eagerly comes at the cost of run-time performance~\cite{Marcinkiewicz2021}. Thus, hybridization approaches~\cite{Moldovan2019,Facebook2019,Apache2021b}
that execute imperative DL programs as graphs at run-time
have been integrated into mainstream DL frameworks. For example, \cref{lst:model} portrays \citetitle{Abadi2016} imperative (OO) DL code representing a modestly-sized model for classifying images. On line~\ref{lne:tfunc}, \citetitle{Moldovan2019}~\cite{Moldovan2019} is used to potentially improve performance by decorating the model's \pyi{call()} method with \TFF, possibly providing optional yet influential decorator arguments. At run-time, \pyi{call()}'s execution will be ``traced'' and an equivalent graph will be generated~\cite{Google2021b}. %
In this case, a speedup ($\nicefrac{\mathit{run time}_{\mathit{old}}}{\mathit{run time}_{\mathit{new}}}$) of $\sim$\num{9.22}, averaged over five runs, ensues~\cite{Khatchadourian2021}.

As noted in \cref{sec:intro}, while promising, hybridization presents unique challenges~\cite{Jeong2019,Google2021b} in ensuring that programs run reliably \emph{and} efficiently. If used incorrectly, hybridization may yield programs that result in unexpected run-time behavior. Decorating the right functions, supplying the correct decorator arguments, using the appropriate API, and properly structuring imperative DL code so that it is amenable to graph execution can be daunting, especially for developers (data scientists) lacking SE expertise.

\textit{Python Side-effects.}\label{sec:motiv:side}
\begin{listing}
    \begin{minipage}[t]{0.48\linewidth}
	\begin{pythoncode}
		@tf.function|\label{lne:func}|
		def f(x): |\label{lne:f}|
			print("Input: ", x) |\label{lne:output}|
		f(1)|\label{lne:f1}|
		f(1)|\label{lne:f2}|
		f(2)|\label{lne:f3}|
	\end{pythoncode}
    \end{minipage}\hfill
    \begin{minipage}[t]{0.48\linewidth}
	Output (expecting \pyc{1}, \pyc{1}, \pyc{2}):
	\begin{pyconcode}
		Input: 1
		Input: 2
	\end{pyconcode}
    \end{minipage}
    \caption{Imperative \citetitle{Abadi2016} code with Python side-effects~\cite{Google2021b}.}\label{lst:se}
\end{listing}
Side-effect producing, native Python statements, e.g., printing, list appending, global variable mutation~\cite{Google2021b}, are problematic for \TFF*-decorated functions.\footnote{Herein, ``\mintinline[fontsize=auto]{python}{tf.function}-decorated'' functions will be referred to as ``\mintinline[fontsize=auto]{python}{tf.function}s.''}
Because they are traced, a function's behavior is ``etched'' into its corresponding graph and thus can have unexpectant results, executing side-effects multiple times or not at all. Side-effects occur when \TFF*s are called the first time; subsequent calls with similar arguments
execute the graph instead. For example, on line~\ref{lne:output} of \cref{lst:se}, \pyi{f()}
outputs
\pyi{x}. On line~\ref{lne:func}, \pyi{f()} is decorated with \TFF, which migrates it to a graph at run-time. Then, \pyi{f()} is invoked three times, the first two with the argument \pyi{1} and the last with \pyi{2}. In the
output
on the right, the first invocation of \pyi{f()} on line~\ref{lne:f1} results in a graph being built (through tracing) that---due to a similar argument---is later used on line~\ref{lne:f2}. Consequently, the side-effecting code on line~\ref{lne:output} is \emph{not} exercised. In contrast, line~\ref{lne:output} \emph{is} exercised as a result of the call on line~\ref{lne:f3} due to a different argument being supplied.

\begin{listing}
    \begin{minipage}[t]{0.48\linewidth}
	\begin{pythoncode}
		class Model(tf.Module):
			def __init__(self):
				self.v = tf.Variable(0)
				self.counter = 0

			@tf.function
			def __call__(self):
				if self.counter == 0:
					self.counter += 1
					self.v.assign_add(1)|\label{lne:vinc}|
				return self.v
	\end{pythoncode}
    \end{minipage}\hfill
    \begin{minipage}[t]{0.45\linewidth}
	\begin{pythoncode*}{firstnumber=last}
		m = Model()
		for n in range(3):
			print(m().numpy())|\label{lne:modelinv}|
	\end{pythoncode*}
	\vspace{1em}
	Output (expecting \pyc{1}, \pyc{1}, \pyc{1}):
	\begin{pyconcode}
		1
		2
		3
	\end{pyconcode}
    \end{minipage}
    \caption{Imperative \citetitle{Abadi2016} code using a counter~\cite{Google2021b}.}\label{lst:counter}
\end{listing}

Although \cref{lst:se} is simple, unexpected behavior can generally be difficult to notice. Consider \cref{lst:counter}, where a model uses a \pyi{counter} to safeguard a variable incrementation. The initial value of \pyi{counter}, however, is captured during tracing upon the first model invocation (line~\ref{lne:modelinv}). The overall effect is that the value of
\pyi{v} is incremented \emph{unconditionally} (line~\ref{lne:vinc}) each time the model is invoked. Such problems are common in migrating deferred-execution--style DL code (e.g., \cref{lst:def}) to an imperative style (e.g., \cref{lst:model}). Worse yet, developers only realize such errors after observing suspicious numerical results or significantly lower performance than expected (e.g., when guarded operations are costly)~\cite{Google2021b}. %

\textit{When To Use Hybridization?}\label{sec:motiv:when}
Besides ensuring that DL code is amenable hybridization~\cite{Berkeley2020}, developers must also know \emph{when} and \emph{where} to use it to avoid performance bottlenecks and other undesired behavior. For example, confusion exists
on how often \TFF\ should be applied~\cite{SEI2020}, and calling \TFF*s recursively
could cause infinite loops~\cite{Google2021b}.
Even if a recursion seems to work,
the \TFF*\ will be traced \emph{multiple} times (``retracing''), potentially impacting performance. Also, using \TFF\ on small computations can be dominated by graph creation overhead~\cite{
Google2022}.

\textit{Using Hybridization Parameters.}\label{sec:motiv:args}
\begin{listing}
    \begin{minipage}[t]{0.585\linewidth}
	\begin{pythoncode}
		model = SequentialModel()
		res1 = model(tf.constant([1, 2, 3]))
		res2 = model(tf.constant([1, 2, 3, 4, 5]))
	\end{pythoncode}
    \end{minipage}\hfill
    \begin{minipage}[t]{0.415\linewidth}
	\vspace{-1.25em}
	\begin{pyconcode}
		WARNING: 5 of the last 5 calls
			triggered ... retracing.
			Tracing is expensive.
	\end{pyconcode}
    \end{minipage}
    \caption{DL model (\cref{lst:model}) client code using varying datasets~\cite{Google2021b}.}\label{lst:modelclient}
\end{listing}
Decorating the \emph{correct} function but with \emph{incorrect} decorator arguments may result in performance degradation. For instance, retracing
helps ensure that the correct graphs are generated for each set of inputs; however, excessive retracing may cause code to run more slowly had \TFF*\ \emph{not} been used~\cite{Google2021b,Roy2021,Yamada2020}. \Cref{lst:modelclient} depicts code that invokes the model declared in \cref{lst:model} multiple times using different (hypothetical) datasets, producing the warning on the right. To limit retracing, an \pyi{input_signature} can be specified on line~\ref{lne:tfunc}, \cref{lst:model} as follows:

\begin{pythoncode*}{linenos=none}
	@tf.function(input_signature=(tf.TensorSpec(shape=[None], dtype=tf.int32),))
\end{pythoncode*}

\noindent A \pyi{[None]} dimension in the \pyi{tf.TensorSpec} allows for flexibility in trace (graph) reuse. Since tensors are matched
on their shape,
a \pyi{None}
wild card allows \TFF*s to reuse traces for variably-sized input---occurring when sequences or images are of different lengths or sizes, respectively. Since each call no longer produces a trace, the warning disappears---averting
any performance bottlenecks.

These simplified examples demonstrate that effectively using hybridization
is not always straight-forward, potentially requiring complex analyses and a thorough understanding of API intricacies---a compounding problem in more extensive programs.
As imperative DL programming becomes more widespread, statistical insight into how such programs are best written efficiently and how to avoid common bugs would be extremely valuable to developers.

\section{Methodology}

\textit{Subjects.}\label{sec:subj}
\begin{table}
    \centering
    \caption{Studied subjects.}\label{tab:subj}
    \footnotesize
    \begin{threeparttable}
	\begin{tabular}{@{}lllllll@{}}
	    \toprule
	    & subj & KLOC & studied periods & cmts/iss & kws & exe \\ \midrule
	    fixes & 122 & 10,879 & 2015-11-06 to 2021-01-14 & 199,140  & 470 & 470 \\ %
	    reports & 167 & 17,378 & 2012-05-07 to 2021-08-11 & 237,232 & 704 & 446 \\ %
	    \midrule
	    Total & 250\tnote{*} & 19,677\tnote{*} & 2012-05-07 to 2021-08-11 & 436,372 & 1,174 & 916 \\ \bottomrule %
	\end{tabular}
	\begin{tablenotes}
	    \item[*] Represents unique totals due to subject overlap between the study portions.
	\end{tablenotes}
    \end{threeparttable}
\end{table}
We
examined
Git commit changesets (code patches; row \textbf{fixes}, \cref{tab:subj}) representing bug fixes involving \TFF*\ and
\GH\ issues (row \textbf{reports}) mentioning \TFF*.
Our study encompassed \NumProjects\
open-source DL
systems (column \textbf{subj}), comprising $\sim$\MLOC\ million lines of source code (column \textbf{KLOC}), \NumCommits\ Git commits (column \textbf{cmts} for \textbf{commits}), \NumIssues\ \GH\ issues (column \textbf{iss} for bug \textbf{reports}), and \NumYears\ years of combined project history, averaging \AvgYears\ years per subject.
Subject details may be found in our dataset~\cite{Velez2022};
subjects sources are publicly available on \GH. %
While we focus \TFF* client usages, we include
\citetitle{Abadi2016}
as
developers often file \GH\ issues against it to discuss \TFF* usage challenges and potential bugs. Subjects include those used in previous studies~\cite{Dilhara2021,Jebnoun2020,Zhang2018,Islam2019,Islam2020a,Liu2020,Chen2021,Zhang2019,Humbatova2020} and
appearing in data science-specific datasets~\cite{Biswas2019}. To determine if a project represents a DL system, i.e., one with at least one DL module, we searched repositories for specific keywords, e.g., ``keras,'' ``layer,'' ``net,'' ``neural network,'' ``deep learning.'' The keywords have also been used in related work~\cite{Jebnoun2020} for a similar purpose; the keywords were only used to ensure that subjects were DL systems, not for finding hybridization bugs. We then verified the code to ensure that the keywords represented DL contexts.

For
changesets (bug fixes),
subject criteria consists of having at least one commit whose changeset contains \TFF*. For
issues,
subjects must have at least one \GH\ issue mentioning ``tf.function.''
Subjects
were mostly written in Python, which is popular for DL~\cite{BenBraiek2018}.
While the subjects include popular open-source repositories from well-known and reputable organizations, e.g., Apache~\cite{ASF2021}, Apple~\cite{Apple2021}, Google~\cite{Google2021a}, NVIDIA~\cite{NVIDIACorporation2021},
they also include
lesser-known repositories to understand hybridization challenges facing the DL community-at-large.
Furthermore, hybridization is relatively new---\TFF* was released on \TFFReleaseDate.

\textit{Mining.}\label{sec:meth:mine}
To find changesets (patches) representing hybridization bug fixes, we mined repositories for commits referencing \TFF*\ using \citetitle{Casalnuovo2017}~\cite{Casalnuovo2017}, a tool for
classifying Git commits used by previous work~\cite{Tian2017,Khatchadourian2020,Tang2022,Babii2021}.
Row \textbf{fixes}, column \textbf{kws} of \cref{tab:subj} is the
commits containing \TFF* in their changesets.
We manually examined all \NumCommitsManuallyExamined\ commits, portrayed by row \textbf{fixes}, column \textbf{exe}.
To find issues related to hybridization, we mined repositories for \GH\ issues mentioning ``tf.function'' by first filtering out issues containing only irrelevant
discussion (e.g., ``social conversation'') using a pre-trained classification model~\cite{Arya2019} used by previous work~\cite{Zhou2021,Wang2020,Pan2021}. We then invoked the \citetitle{GitHub2021}~\cite{GitHub2021} to select (open and closed) issues that included ``tf.function'' using several different criteria, e.g., ``best match,'' ``most commented.'' To reduce false positives, since the API ignores punctuation, we further filtered the results to ensure that they included the period. Row \textbf{reports}, column \textbf{kws} of \cref{tab:subj} is the
issues\footnote{Also includes pull (patch) requests as these are treated similarly in \GH.} containing ``tf.function'' in either their title or body (description and conversations).
We
randomly selected a subset of these
to examine manually (details below), portrayed by row \textbf{reports}, column \textbf{exe}. The aforementioned tools~\cite{Casalnuovo2017,Arya2019,GitHub2021} were only used to narrow the search space and not for classification, which was done manually. The \GH\ search was performed in a (standard) manor consistent with previous work.

\textit{Identification.}
We used a \citetitle{Casalnuovo2017} feature that leverages heuristics based on
log messages to identify bug fix commits. Natural language processing (NLP) is internally used by \citetitle{Casalnuovo2017} to determine the commits that fall into this category. Doing so helps us to focus on likely bug fix commits for further manual examination. Random matching issues---with ones containing code
being favored---were chosen for manual inspection.
Next, the authors manually examined the commits and issues to ascertain if they indeed relate to hybridization bugs. Two authors are SE and PL professors with extensive expertise in software evolution, system performance, and empirical SE\@. Another author is a data mining and ML professor with substantial proficiency in AI and SE\@. Three authors have several years of industrial SE experience.

Although the researchers did not converse during the initial identification and classification process to avoid bias, this mix of expertise is effective in studying SE tasks in DL systems. The researchers convened regularly during the study, as well as at the end for finalization, to solidify the results. Cohen's Kappa coefficients~\cite{Viera2005} for identification and classification were \num{0.80} and \num{0.57}, respectively.\footnote{Moderate agreement is expected; the team has mixed ML/SE expertise.} As the authors did not always have detailed knowledge of the particular systems, only changes where a bug fix was extremely likely were marked as such. The authors also used commit comments and referenced bug databases to ascertain whether a change was a bug fix.
\GH\ issues tags were also considered.

\textit{Classification.}\label{sec:meth:class} %
For commits, once bug fixes were identified, the authors studied the code changes to determine the category of bug fixes and whether the category relates to hybridization. For issues, the authors examined issue descriptions and discussions, paying
attention to the \TFF*\ challenges being described and their possible solutions and workarounds. Particular attention was paid to code snippets. No scripts were involved in the classification---only manual examination. Categories were then formed into a hierarchy, in part by using the \citetitle{Abadi2016} documentation~\cite{Google2021b}.
On several occasions, developers were contacted for clarification using the \GH\ line comment mechanism and via email.

\section{Results}\label{sec:res}

This section answers \cref{rq:bugs} by summarizing our results, noting trends, exceptions, and unexpected outcomes.
Contrarily, \Cref{sec:discuss} consolidates, comments on, and connects the main findings.
Related discussion in \cref{sec:discuss} is referenced where appropriate.

\subsection{Quantitative Analysis}\label{sec:res:quant}

From the \NumCommitsManuallyExamined\ commits and \NumIssuesManuallyExamined\ \GH\ issues (totaling \NumManuallyExamined) manually examined (column \textbf{exe}, \cref{tab:subj}), we found \NumBugs\ and \NumDifficulties\ (totaling \NumChallenges) \TFF*\
bug fixes and developer challenges depicted in columns \textbf{cmts} (commits) and \textbf{iss} (\GH\ issues) of \cref{tab:cats}, respectively. Finding these bugs and understanding their relevance required a significant amount of manual labor that may not be feasible in more large-scale, automated studies. Python, being a dynamic language, can be difficult to analyze, particularly w.r.t.~inheritance relationships; subclassing Keras models is a common way to write imperative DL code in \citetitle{Abadi2016} (cf.~line~\ref{lne:subclass}, \cref{lst:model}). Furthermore, our number of findings (\NumChallenges) is comparable with previous studies involving manual inspection (e.g., \citet{Tang2021} found \NumRefactTang, \citet{Zhang2018} found \NumBugsZhang, \citet{Khatchadourian2020} found \NumBugsKhatchadourian). Nevertheless, as \TFF*\ becomes more popular,
we expect its usage and number of related bugs to grow.

\subsubsection{Problem Categories}\label{sec:res:probs}

\begin{figure*}[t]
    \includegraphics[width=0.8445\linewidth,center]{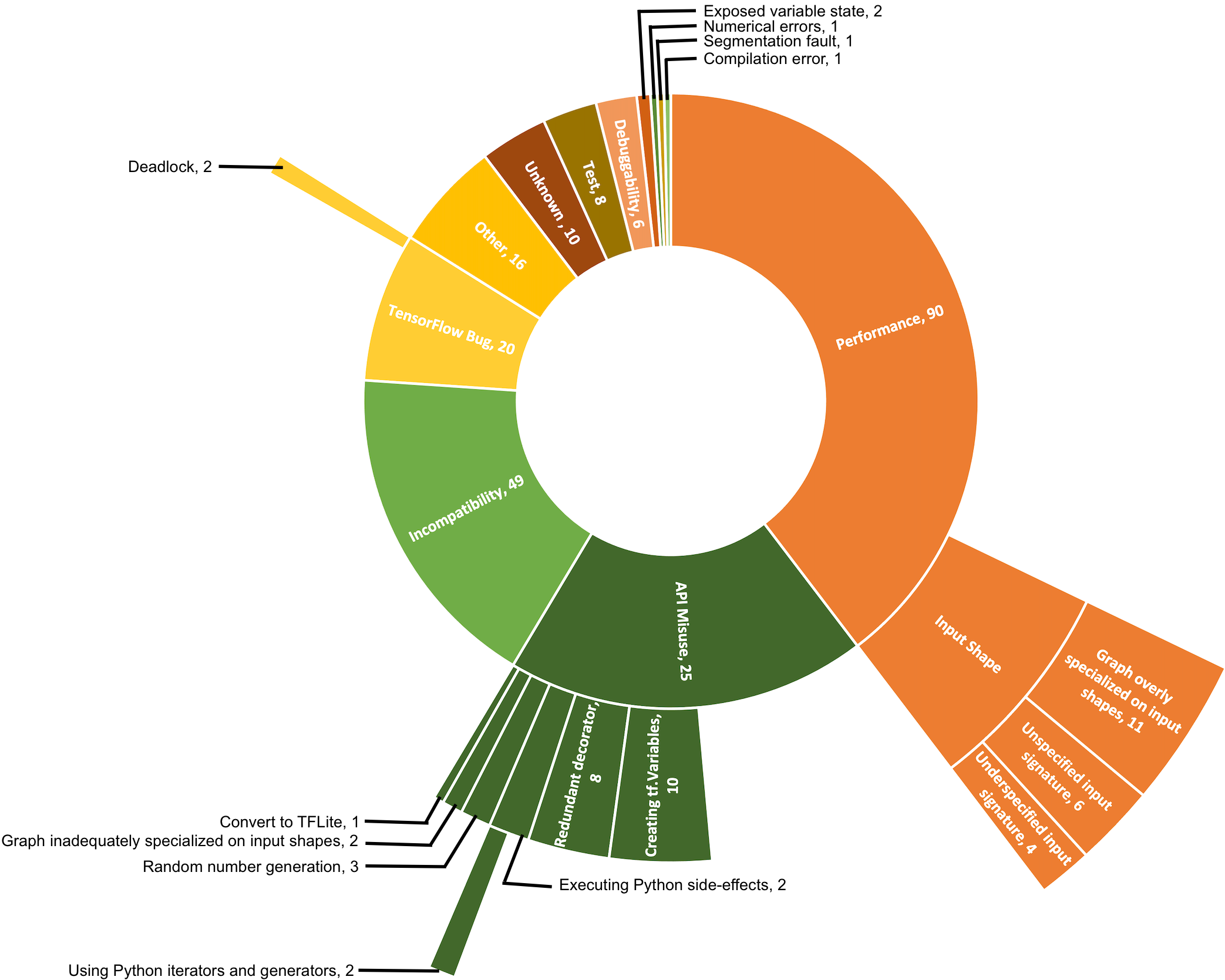}
    \caption{Discovered problem categories (hierarchical).}\label{fig:cats}
    \Description[Problem category sunburst.]{Sunburst diagram displaying a problem category hierarchy.}
\end{figure*}

\begin{table}
    \centering
    \footnotesize
    \caption{Discovered top-level problem categories.}\label{tab:cats}
    \begin{tabular}{@{}llllllllllll@{}}
	\toprule
	problem & abbr & cmts & iss & total \\ \midrule
	Performance & PRF & 74 & 37 & 111 \\
	API misuse & APM & 23 & 30 & 53 \\
	Incompatibility & INC & 16 & 33 & 49 \\
	TensorFlow bug & TFB & 4 & 18 & 22 \\
	Other & OTH & 14 & 2 & 16 \\
	Unknown & UKN & 10 & 0 & 10 \\
	Test & TST & 8 & 0 & 8 \\
	Debuggability & DBG & 4 & 2 & 6 \\
	Exposed variable state & EVS & 1 & 1 & 2 \\
	Compilation error & CMP & 1 & 0 & 1 \\
	Numerical errors & NME & 1 & 0 & 1 \\
	Segmentation fault & SEG & 1 & 0 & 1 \\
	\midrule
	Total &  & 157 & 123 & 280 \\ \bottomrule
    \end{tabular}
\end{table}

We group bug fixes and \GH\ issues into common problem categories, shown in \cref{fig:cats,tab:cats}
(column \textbf{abbr}
is the category abbreviation). The former includes combined data (commits and issues), while the latter separates the two. \Cref{fig:cats} presents a hierarchical categorization---with varying levels of detail---of the \NumChallenges\ discovered \TFF*-related challenges
in our
subjects. Challenges are represented by their problem category name and are followed by their counts. Categories without instances are
\emph{abstract}, i.e., they \emph{only} group together other categories. \Cref{tab:cats} portrays a nonhierarchical, top-level view of \cref{fig:cats};
the innermost (top) layers of \cref{fig:cats} represent the rows of \cref{tab:cats}.

Challenges are grouped into several (top-level) problem categories. Categories include performance (PRF, \num{90}; further discussed later),
API misuse (APM, \num{25}; further discussed later),
and incompatibility between execution modes, i.e., eager and deferred, where \TFF*\ is used in a context not amenable to graph conversion (INC, \num{48}; further
discussed later).
An example of the latter is where particular loss functions cannot be used in graph mode or there is an \citetitle{Moldovan2019} limitation that prevents graph conversion.
Other problem categories include dealing with or working around open bugs related to \TFF*\ in \citetitle{Abadi2016} (TFB, \num{20}; further discussed later)
and ``other'' (OTH, \num{16}), which involves syntactic corrections, general cleanup, and refactorings---a category similar to that used by previous work~\cite{Tian2017,Khatchadourian2020}. ``Unknown'' (UKN, \num{10}) represents situations where the problem category was indeterminable without further domain knowledge or developer input. Only \UnknownCategoriesPercentage\ of problems had unknown categories. Code changes involving \TFF*\ appearing in tests were categorized as ``Test'' (TST, \num{8}).

\textit{Debuggability. }\label{sec:res:quant:dbg}
Debuggability (DBG, \num{6}) represent situations where using \TFF*\ to improve performance of DL code may, in turn, reduce a developer's ability to easily debug it. ``In general, debugging code is easier in eager mode than inside \TFF*''~\cite{Google2021b}.
In such situations, developers may not understand that using \TFF*\ is the reason why they are not able to debug their code, e.g., intermediate variable values may be missing. Or, \TFF*\ may \emph{temporarily} be removed (via a commit) to facilitate debugging, but developers inadvertently neglect to replace it (cf.~\cref{sec:res:qual:dbg}). This latter situation is unfortunate as, to assist in the debugging process, a flag
can be used to globally (temporarily) toggle \TFF*~\cite{Google2021b}. %

\textit{Other Categories. }\label{sec:res:quant:others}
Other (top-level) categories were more minor in terms of their counts, yet have potentially significant consequences. For example,
exposed variable state (EVS, \num{2}) occurs when saving (exposed) program state (variables) is problematic during \TFF*\ conversion at run-time, e.g., variables becoming undefined~\cite{Huang2020}. Numerical errors (NME, \num{1}) involve possible numeric overflow. %
Autograph compilation errors (CMP, \num{1}) surface when \TFF*s are compiled and subsequently result in compilation errors. This problem may arise when certain dynamic Python features, e.g., lexical scoping, are utilized (cf.~\cref{sec:res:qual:cmp}). Segmentation fault (SEG, \num{1}) is when using \TFF*\ causes a program crash. While compilation and numerical errors and segmentation faults may be considered symptoms, we focus on \TFF*\ \emph{client} usage; these categories represent problems from a \emph{client} perspective. Their underlying causes are bugs within the framework.

\textit{Performance.}\label{sec:res:perf} %
\begin{table}[t]
    \centering
    \footnotesize
    \caption{Performance fixes.}\label{tab:perffixes} %
    \begin{tabular}{@{}ll@{}}
	\toprule
	fix category & count \\ \midrule
	Add \mintinline[fontsize=auto]{python}{tf.function} decorator & 61 \\
	Change \mintinline[fontsize=auto]{python}{tf.function} argument & 20 \\
	Add \mintinline[fontsize=auto]{python}{input_signature} argument to \mintinline[fontsize=auto]{python}{tf.function} & 9 \\
	Remove \mintinline[fontsize=auto]{python}{tf.function} decorator & 8 \\
	Upgrade to new library version & 4 \\
	Relocate \mintinline[fontsize=auto]{python}{tf.function} (use on different function) & 5 \\
	Re-add \mintinline[fontsize=auto]{python}{tf.function} decorator & 2 \\
	Unsolved (open) & 2 \\
	Total & 111 \\ \bottomrule
    \end{tabular}
\end{table}
As the main purpose to hybridization is to improve the performance of imperative style DL code by building a bridge to graph-based execution, it was not surprising that performance---at \PerformancePercentage\ (\PerformanceFraction)---was the largest category:

\finding{At \PerformancePercentage\ (\PerformanceFraction), performance was the largest problem category encompassing \TFF*\ usage.\label{fnd:prf}}

\noindent Performance problems represent a spectrum of situations, stemming from using \TFF*\ to solve a DL code performance bug to not observing the expected speedup from using \TFF*\ to exhibiting \emph{worse} performance that \emph{not} using \TFF*. \Cref{tab:perffixes} portrays the various fixes used to solve performance problems. %
Though the majority of times it was used to enhance performance of imperative DL code, we found that in \RemovedTFFPercentage\ (\RemovedTFFFraction) of cases, \TFF* was \emph{removed} to alleviate performance problems:

\finding{Despite intent to improve
performance, \TFF* \emph{caused} performance degradation in \RemovedTFFPercentage\ (\RemovedTFFFraction) of cases.\label{fnd:badperf}}

\noindent Moreover, only \PerfFixedByAddingTFFPercentage\ of imperative DL code performance problems were fixed by \emph{adding} \TFF*. Thus, the remaining \PerfFixedOtherPercentage\ of cases were due to \emph{existing} hybridization:

\finding{Only \PerfFixedByAddingTFFPercentage\ (\PerfFixedByAddingTFFFraction) of imperative DL code performance problems were fixed by \emph{adding} \TFF*. The remaining \PerfFixedOtherPercentage\ were due to \emph{using} \TFF*.\label{fnd:prffixes}}

\noindent In fact, \PerfFixedByArgsPercentage\ of performances fixes involved altering \TFF* arguments:

\finding{Performance fixes entailed altering developer-supplied \pythoninline{tf.function(...)}  arguments at a rate of \PerfFixedByArgsPercentage.\label{fnd:perfargfix}}

Performance problems are further categorized into those related to ``input shapes,'' which make up \InputShapePercentageOfPerformanceProblems\ of all such problems:

\finding{Performance problems involved incorrect input tensor shape specifications at a rate of \InputShapePercentageOfPerformanceProblems.\label{fnd:shapes}}

\noindent Tensors are heavily used DL programs, and accurately matching tensor shapes (dimensions) is often required to write reliable DL code. In hybridization, since \TFF*s are being traced and thus converted into graphs, the underlying framework (by default) attempts to build specialized graphs for each kind of input. However, when tensors are involved, graphs may be specialized to particular input shapes, creating a situation where function retracing is excessive. Retracing can lead to significant performance degradation~\cite{Google2022}.

To curb this problem, an (\pythoninline{input_signature}) argument may be supplied to \TFF*\ that specifies an \emph{expected} range of shapes.
In effect, developers
provide contextual information to the framework about how \TFF*s will be used. For instance, setting
\pythoninline{experimental_relax_shapes}
to \pythoninline{True} may cause \TFF*s to generate fewer graphs that are less specialized on input shapes. However, this
may not match reality, especially when dealing with dynamic shapes. As such, we further divide ``input shape'' challenges:

\begin{description}
    \item[Graph \emph{overly} specified on input shapes (11)] Generated graphs are too specific for the context where a \TFF* is being used, which can occur when either:
	\begin{enumerate}[(i)]
	    \item \pythoninline{experimental_relax_shapes} is incorrectly set to \pyi{False}.
	    \item \pyi{input_signature} is unnecessarily specified. Either it should be either removed or set to \pyi{None} (the default).
	\end{enumerate}
    \item[\emph{Under}specified input signature (4)] The \pyi{input_signature} parameter
	lacks proper arguments to avoid excessive retracing.
    \item[\emph{Un}specified input signature (6)] The \pyi{input_signature} is missing in contexts that are advantageous to graph specialization.
\end{description}

\textit{API Misuse.}\label{sec:res:api}
\begin{table}
    \centering
    \footnotesize
    \caption{API misuse causes.}\label{tab:apicauses} %
    \begin{tabular}{@{}ll@{}}
	\toprule
	cause & count \\ \midrule
	API confusion & 20 \\
	Use of graph mode & 14 \\
	Decorated outer function calls unnecessarily decorated inner function & 8 \\
	Incorrect \mintinline[fontsize=auto]{python}{tf.function} argument & 7 \\
	Use of eager mode & 2 \\
	Lost variable state due to graph conversion & 1 \\
	Lack of static shape specifications & 1 \\
	Total &  53 \\ \bottomrule
    \end{tabular}
\end{table}
API Misuse---the second largest problem category at \APIMisusePercentage---involves situations where \TFF*\ is not used in a way recommended by the API documentation:

\finding{At \APIMisusePercentage, API misuse---using \TFF*\ inconsistent to documentation---was the \nth{2} largest
category.\label{fnd:apm}}

\noindent Misusing APIs typically results in either run-time errors or unexpected behavior. Violating DL API constraints may lead to crashes and poor performance~\cite{Islam2019,Islam2020a}. In high-level, e.g., imperative DL, code,
bugs are commonly due to misunderstandings of the guarantees offered and obligations imposed by increasingly layered software, e.g., those written against the \citetitle{Abadi2016} API~\cite{Lagouvardos2020}.
\citetitle{Abadi2016} documentation contains a prominent sections regarding \TFF* and \citetitle{Moldovan2019} usage constraints and limitations.
If such constraints, e.g., w.r.t.~control-flow, side-effects, global variables, are violated, \citetitle{Moldovan2019} will not
properly generate graphs from Python code. Despite the vast
documentation, at \APIConfusionPercentage, API confusion was the largest cause of API misuse (\cref{tab:apicauses}):

\finding{API misuse was caused by developers not understanding
hybridization APIs
at a rate of \APIConfusionPercentage\ (\APIConfusionFraction).\label{fnd:apmcause}}

Regarding potential category overlap, recall that API misuse is defined above as a violation of intended API usage per the documentation. Consider changing a \TFF*\ argument. Performance degradation can occur when parameter usage is \emph{consistent} with the documentation; it can be a tuning issue, e.g., shape-related. In such a case, according to the earlier definition, shape mismatches would not be considered API misuse as they are dependent on context.

\begin{figure}
    \centering
    \includegraphics[width=\linewidth]{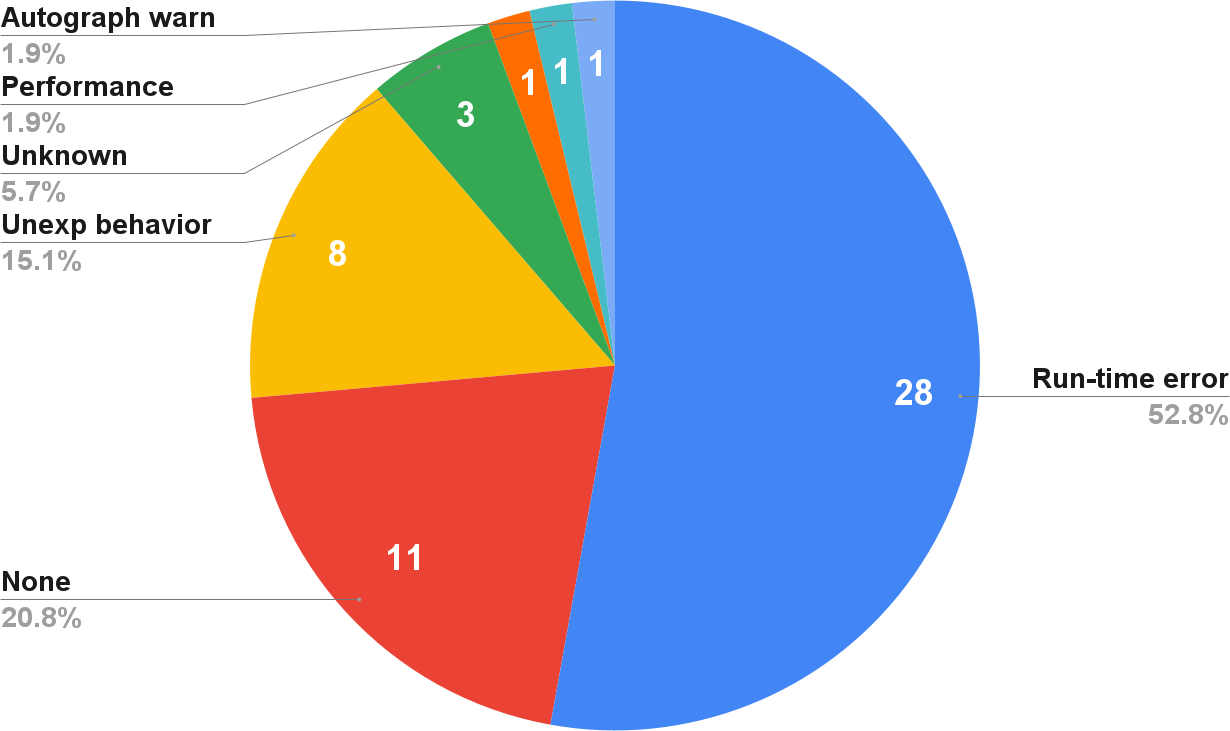}
    \caption{API misuse symptoms.}\label{fig:api_symptoms}
    \Description{Pie chart of API misuse symptoms.}
\end{figure}

We found that the most common way (\APIMisuseFixedByRemovalPercentage\ or \APIMisuseFixedByRemovalFraction) to fix API misuse was to \emph{remove} \TFF. Of these, in \SI{46.67}{\percent} of cases ($\nicefrac{7}{15})$, the problem cause was that \TFF\ was used to decorate an \emph{inner} function %
called by an \emph{already} decorated \emph{outer} function. As \TFF*\ applies to the decorated function \emph{and} all other functions it calls and since the inner function cannot be called from any other function besides the outer function, the inner function decorator is unnecessary and can thus be safely removed~\cite{Google2022}.
However, another \SI{46.67}{\percent} ($\nicefrac{7}{15}$) of cases were caused by API confusion. Thus, in these cases, unfortunately, developers \emph{abandoned} \TFF---along with its potential to enhance performance---due to their confusion over how to use it. Most likely, developers were doing so to avoid run-time errors, which occurred in \SI{62.50}{\percent} ($\nicefrac{5}{8}$) of \TFF*\ removals not caused by unnecessary inner function decoration and \SI{52.83}{\percent} ($\nicefrac{28}{53}$) overall (\cref{fig:api_symptoms}):

\finding{To fix API misuse, \TFF*\ was \emph{removed} \APIMisuseFixedByRemovalPercentage\ of the time. In \SI{46.67}{\percent} of these, hybridization was abandoned due to API confusion, with \SI{62.50}{\percent} causing run-time errors.\label{fnd:apmfixremove}} %

API misuse is further divided into several categories, the largest of which involves creating \pythoninline{tf.Variables} within \TFF*s (10). A \pyi{tf.Variable} represents a tensor whose value is mutable~\cite{Google2021e}. Currently, \TFF*\ only supports singleton
\pyi{tf.Variable}s; creating multiple \pyi{tf.Variable}s within the scope of a \TFF*\ results in a run-time exception~\cite{Google2021b}.
Redundant decoration (8) is where \emph{multiple} functions on a call path are unnecessary decorated with \TFF; all functions called from a \TFF*\ are also automatically migrated to graphs. Accurately approximating such paths statically---especially in the context of a dynamic language such as Python---may be difficult, and there is ample confusion among developers on where to apply \TFF~\cite{SEI2020} (cf.~\cref{sec:motive}). %

Executing Python side-effects (2) refers to the situation
where \TFF*s contain side-effect producing Python statements. As described in \cref{sec:motiv:side}, executing such statements within migrated graphs can have unexpectant results, sometimes executing twice or not all. A specific pattern of side-effects were those involving the use of iterators and generators (2), a common looping mechanism in Python code. Random number generation (RNG, 3) problems occur when developers do not use RNG facilities consistently with the documentation, commonly resulting in unexpected behavior under graph mode. For example, RNG creation inside a \TFF*\ can only happen during the first run of the function~\cite{Google2021o}. Seeding may also not work as expected in graph mode (e.g.,~\cite{Geron2019})---``when [a] global seed is set but [\citetitle{Abadi2016}] operation seeds are not, the sequence of random numbers are the same for each \TFF*''~\cite{Google2021j}. ``Graph \emph{inadequately} specialized on input shapes'' (2) involves an API misuse that is opposite to the ``graph \emph{overly} specified on input shapes'' \emph{performance} problem category described
earlier. Such problems may be fixed by setting \pythoninline{experimental_relax_shapes} to \pyi{False} (the default).
In other words, the shape specification is too \emph{general}, which may result in a situation that is not amenable to graph migration~\cite{Poluektov2021}. For example, an \pyi{input_signature} may be supplied using a wild card shape to improve performance (q.v.~\cref{sec:motiv:args}) but results in a run-time error due to a tensor dimension mismatch~\cite{Google2021b,
Yi2022}. Conversion to TFLite (1) represents problems with an alternate use case of \TFF* to convert a DL model to a portable format. %

\textit{Execution Mode Incompatibility. }\label{sec:res:inc}
At \IncompatabilityPercentage, incompatibility is the third largest problem category:

\finding{Execution mode
    incompatibility, at \IncompatabilityPercentage\ (\IncompatabilityFraction), was the \nth{3} largest problem category, meaning that seamlessly using similar constructs in different modes was problematic.\label{fnd:inc}}

\noindent Developers seemingly struggle with seamlessly using imperative DL program constructs, e.g., particular loss functions, across execution modes. Ideally, developers could toggle between eager and graph execution modes---with \citetitle{Moldovan2019} simply enhancing performance---without making code changes. In other words, incompatibility problems prevents developers from focusing on the correctness of their DL code---thinking of performance as an afterthought. Instead, to use hybridization effectively, developers must be cognizant of its internal structure, i.e., \emph{how} their DL code is being migrated to graphs.
Moreover, developers must (manually) be aware of which constructs are amenable to graph conversion, how best to write code that works in either mode, and how to interact with code that may be executed in a different mode.

\begin{figure}
    \centering
    \includegraphics[width=\linewidth]{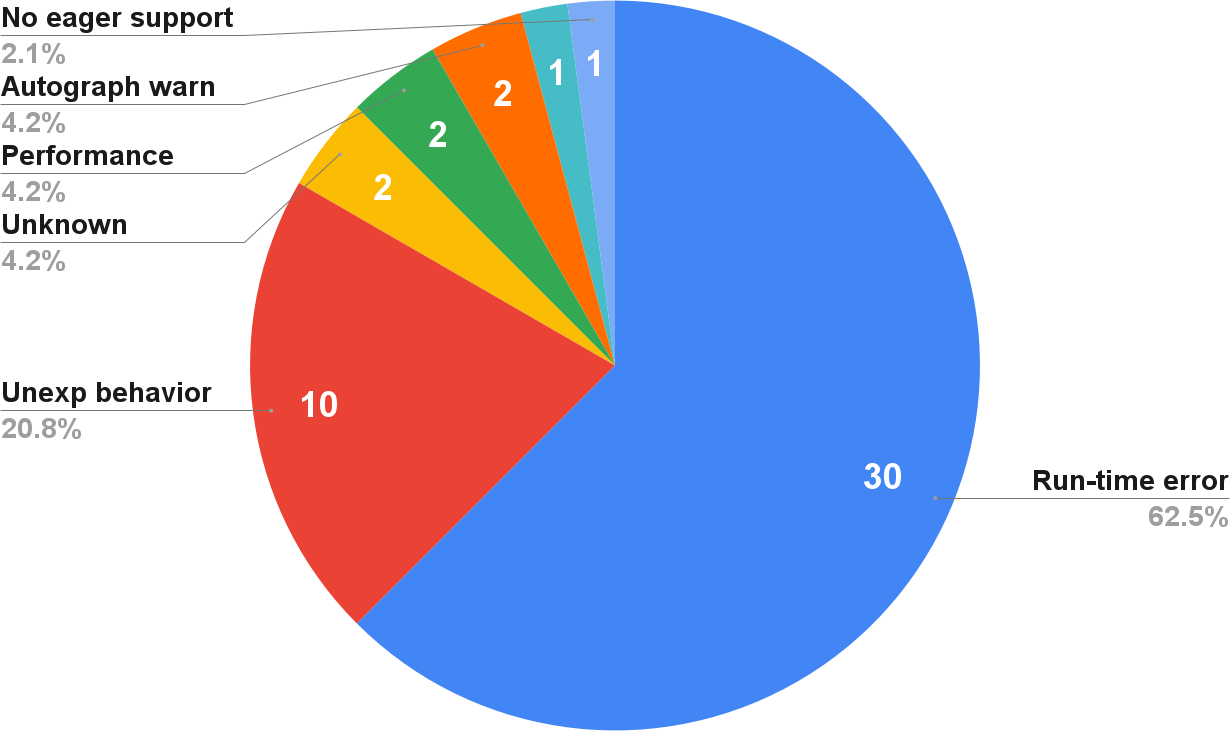}
    \caption{Incompatibility problem symptoms.}\label{fig:inc_symptoms}
    \Description{Pie chart of incompatibility problem symptoms.}
\end{figure}

Execution mode incompatibility problems have dire consequences. As shown in \cref{fig:inc_symptoms},
\RuntimeErrorUnexpectedBehabiorIncompatabilitySymptomPercentage\ of symptoms resulting from incompatibility involve run-time errors or unexpected behavior. Such problems that only occur at run-time are difficult to uncover and, if found, may be found after deployment:

\finding{Incompatibility problems led to run-time errors or unexpected results, which do not surface until \emph{after} running the
code, \RuntimeErrorUnexpectedBehabiorIncompatabilitySymptomPercentage\ of the time.\label{fnd:incsympt}}

\textit{TensorFlow Bugs. }\label{sec:res:tfb}
\citetitle{Abadi2016} bugs (TFB) made up \TensorFlowBugsPercentage\ of bugs:

\finding{\citetitle{Abadi2016} bugs, where developers were offered workarounds or awaited new framework versions, made up \TensorFlowBugsPercentage\ of problems. Of these, \PercentagOfTensorFlowBugsLeadingToDeadlock\ involve deadlocks.\label{fnd:tfb}}

\noindent Such bugs involve dealing with or working around open \citetitle{Abadi2016} bugs related to \TFF*. As hybridization is relatively new, the \TFF*\ API is under active development.
Thus, it was not uncommon for such bugs to be reported to \citetitle{Abadi2016} by filing issues against its \GH\ repository; \TensorFlowBugsAsIssuesPercentage\ of TFBs appear as \GH\ issues (see \cref{tab:cats,fig:cat-compare}). We categorize bugs as TFB if they were in fact real bugs with \citetitle{Abadi2016} that required a workaround---often suggested by \citetitle{Abadi2016} contributors---or a new \citetitle{Abadi2016} library version to solve. If the reported bugs were not resolved to be the result of problems with \citetitle{Abadi2016}, such bugs were not categorized as TFB but perhaps other categories.

TFB is further categorized into deadlock (2). Situations leading to the execution of a
\TFF*\ being deadlocked include using tensors as stopping condition of a recursive \TFF*~\cite{Xu2021}. Deadlock may also occur as a result of other, specific \TFF*\ code patterns---causing the \citetitle{Abadi2016} run time to deadlock. For example, deadlock may occur when calling a \TFF*\ from within a \pyi{tf.py_function}~\cite{Xu2019}, which executes native Python functions as graph operations eagerly~\cite{Google2021m}.

\subsubsection{Commits vs.~\GH\ Issues}\label{sec:res:comvsiss}

\begin{figure}
    \centering
    \includegraphics[width=\linewidth]{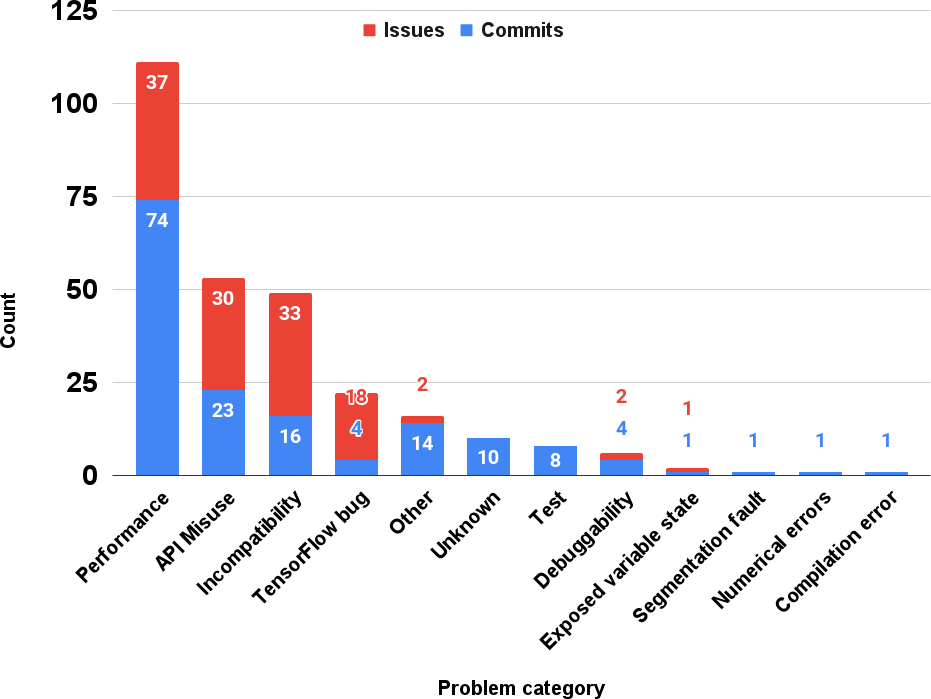}
    \caption{Top-level problem category comparison.}\label{fig:cat-compare}
    \Description[Stacked bar chart.]{A stacked bar chart of top-level problem categories comparing commits to \GH\ issues.}
\end{figure}

\Cref{fig:cat-compare} compares the different sources---commits (bottom/\textcolor{blue}{blue} bars) and \GH\ issues (top/\textcolor{red}{red} bars)---of
problem categories. Performance---the largest problem area---was \nicefrac{2}{3} more likely to appear in \emph{commits} vs.~issues. In contrast, ``incompatibility'' was
\nicefrac{2}{3} more likely to appear in \emph{issues} vs.~commits. Notably, \SI{80}{\percent} of TFB problems were found in \GH\ issues compared to commits. Lastly, all UKN and TST bugs were found in commits. \Cref{sec:discuss:vs} discusses possible reasons for these differences.

\subsection{Qualitative Analysis}\label{sec:res:qual}

This section answers \cref{rq:ex} by highlighting bug patterns with examples, summarizing causes, symptoms, and fixes, and proposing preliminary best practices
and anti-patterns.

\begin{figure}[t]
    \small
    \begin{bestpractices}
	\item Favor \mintinline[fontsize=auto]{python}{@tf.function} on Python functions containing imperative, otherwise eagerly-executed, DL code to improve performance.\label{bp:perf}
	\item If possible, supply an \mintinline[fontsize=auto]{python}{input_signature} argument to \mintinline[fontsize=auto]{python}{tf.function} with the intended shape and types of any input tensors to avert retracing---a practice similar to that of providing type annotations to variables in dynamic languages to assist with type inferencing.\label{bp:ins}
	\item When an operation is deemed incompatible with hybridization, check the documentation to see if additional steps are required to make the imperative DL code more amenable to graph conversion.\label{bp:docs}
	\item Framework limitations may impede performance enhancements. Check
	    for potential workarounds of (unresolved) \citetitle{Abadi2016} bugs.\label{bp:bugs}
	\item Use \mintinline[fontsize=auto]{python}{tf.config.run_functions_eagerly(True)} to \emph{temporarily} disable \mintinline[fontsize=auto]{python}{tf.function} to facilitate debugging.\label{bp:dbg}
    \end{bestpractices}
    \caption{Preliminary hybridization best practices.}\label{fig:bps}
\end{figure}

\begin{figure}[t]
    \small
    \begin{antipatterns}
	\item Hybridizing
	    nested
	    functions may cause performance degradation.
	    Sacrifice some modularity by either hybridizing the
	    top-level
	    function or refactoring the
	    nested function to a
	    top-level
	    function~\cite{Sung2020}.\label{ap:nested}
	\item Since shared variables must be singleton, using \mintinline[fontsize=auto]{python}{tf.Variable}s in \mintinline[fontsize=auto]{python}{tf.function}s, either directly or indirectly, may cause run-time exceptions. Either rewrite the function or do not hybridize it.\label{ap:vars}
	\item Since \mintinline[fontsize=auto]{python}{tf.function}s are compiled, using dynamic language features, e.g., lexical scoping, either directly or indirectly,
	    may lead to run-time exceptions. Avoid such features in \mintinline[fontsize=auto]{python}{tf.function}s where possible.\label{ap:dyn}
    \end{antipatterns}
    \caption{Preliminary hybridization anti-patterns.}\label{fig:aps}
\end{figure}

\subsubsection{Performance}\label{sec:res:qual:prf}

\begin{commit}{\galference, \href{https://github.com/modichirag/galference/commit/af1664e7f3cad8c707dd45941c6a56dc9d408314}{\textinline{af1664e7}}, code/recon_dm-tf2-lbfgs.py, Chirag Modi, , Dec, 25, 2020, bug boxsize=nc, prf}
    \begin{diffcode}
|\label{lne:addtf}|+ @tf.function
  def pm(linear):
  	state = lpt_init(linear, a0=0.1, order=1)
  	final_state = nbody(state,  stages, nc)
  	tfinal_field = cic_paint(tf.zeros_like(linear), final_state[0])
  	return tfinal_field
    \end{diffcode}
\end{commit}

In \cref{lst:prf}, \pyi{pm()} is decorated with \TFF\ (line~\ref{lne:addtf}). Using \TFF*\
``ensure[s] that the graph for [a] function is compiled once and not every time it is called, thus gaining in speed and performance''~\cite{Modi2021}, leading to \cref{bp:perf}, \cref{fig:bps}.

While hybridization can enhance the performance of their imperative, otherwise eagerly-executed, DL code, we found that developers struggled to use it correctly. Some distrusted it, stating, e.g., that ``it does far too much hidden magic''~\cite{Roberts2020}. Others~\cite{Mondal2020} struggled with uncontrolled retracing (q.v.~\cref{sec:motiv:when,sec:res:perf}), which actually results in worse performance---speedup of \num{0.13} in this case---by using \TFF*\ than not using it: ``\pyi{tfa.image.equalize()} uses an internal \pyi{scale_channel()} function[,] which triggers excessive retracing \ldots.''
The problem is related to hybridizing \emph{inner} functions:  ``I \ldots\ tried using \TFF\ at the \pyi{scale_channel()} and \ldots\ \pyi{equalize_image()} level[s], but the further I moved it `inside,' the slower \pyi{equalize()} became''~\cite{Mondal2020a}. The fix involved ``using \TFF\ at the top-level of \pyi{equalize()}, which made it run $\sim$\SI{25}{\percent}--\SI{40}{\percent} faster \ldots.'' The root cause is that, ``using [embedded] functions ([i.e.,] defining functions inside function) will retrace the graph multiple times [as] the[ir] scope is not [publicly] visible, and the graphs cannot be cached''~\cite{Sung2020}. As a modularity mechanism, embedding (nesting) function definitions is a common idiom in Python,
yet,
currently, \citetitle{Abadi2016} documentation does not mention this problem. Developers are left to consider the internals of \citetitle{Moldovan2019} in writing performant imperative DL code, leading to \cref{ap:nested}, \cref{fig:aps}.

\textit{Input Signatures.}\label{sec:res:quant:ins}
\begin{commit}{\DDPG, \href{https://github.com/samuelmat19/DDPG-tf2/commit/02a3f297dae2f2a0643e0799887e1f253ee46c2a}{\textinline{02a3f297}}, model.py, Samuel Matthew Koesnadi, samuelmat19@gmail.com, Sep, 21, 2020, Fixed all\ldots this should work, ins}
    \begin{diffcode}
- @tf.function
|\label{lne:ins_start}|+ @tf.function(input_signature=[
+ 	tf.TensorSpec(shape=(None, self.num_states), dtype=tf.float32),
+ 	tf.TensorSpec(shape=(None, self.num_actions), dtype=tf.float32),
+ 	tf.TensorSpec(shape=(None, 1), dtype=tf.float32),
|\label{lne:ins_end}|+ 	tf.TensorSpec(shape=(None, self.num_states), dtype=tf.float32),])
  def update_weights(s, a, r, sn): # ...
    \end{diffcode}
\end{commit}
Arguments to \TFF*\pyi{()}, particularly involving input tensor shapes, may also influence performance (q.v.~\cref{sec:motiv:args}). \Cref{lst:ins} portrays an underspecified input signature
(q.v.~\cref{sec:res:perf})---one of the most used \TFF*\ parameters that we observed. On lines~\ref{lne:ins_start}--\ref{lne:ins_end}, a performance regression was fixed by adding an \pyi{input_signature}
to a
weight distribution \TFF*\ to ``make sure it does not recreate graph, which will slow down training significantly''~\cite{Koesnadi2021}. The sequence of \pythoninline{tf.TensorSpec}s specifies the intended tensor shapes and data types (\pyi{dtype}s) that will be supplied to \pyi{update_weights()}. Otherwise, a separate (concrete) function (graph) is instantiated for each \emph{inferred} input signature, which
may result in retracing, leading to \cref{bp:ins}, \cref{fig:bps}.

\subsubsection{Compilation Errors}\label{sec:res:qual:cmp}

\begin{commit}{\GPflow, \href{https://github.com/GPflow/GPflow/pull/1418/commits/b65848a22e258dcff566bd5516a431eefae860ee}{\textinline{b65848a2}}, gpflow/quadrature.py, Vincent Dutordoir, dutordoirv@gmail.com, Apr, 8, 2020, fix compilation issue\ldots, cmp}
    \begin{diffcode}
  def ndiagquad(funcs, H: int, Fmu, Fvar, logspace: bool = False, **Ys):
  	# Computes N Gaussian expectation integrals of one or more functions ...
|\label{lne:uni_start}|- 	def unify(f_list): # Stack a list of means/vars into a full block.
- 		return tf.reshape(tf.concat([tf.reshape(f, (-1, 1)) for f in f_list],
|\label{lne:uni_end}|- 			axis=1), (-1, 1, Din))
  	if isinstance(Fmu, (tuple, list)):
|\label{lne:din_def}|  		Din = len(Fmu)
|\label{lne:uni_new}|+ 		def unify(f_list): # Stack a list of means/vars into a full block.
+ 			return tf.reshape(tf.concat([tf.reshape(f, (-1, 1)) for f in f_list],
+ 				axis=1), (-1, 1, Din))
|\label{lne:uni_call}|  		Fmu, Fvar = map(unify, [Fmu, Fvar])  # both [N, 1, Din]
    \end{diffcode}
\end{commit}

Consider \pyi{unify()} originally defined on lines~\ref{lne:uni_start}--\ref{lne:uni_end}, \cref{lst:cmp} that accesses \pyi{Din} on line~\ref{lne:uni_end}. This variable, however, it is defined \emph{after} the function definition on line~\ref{lne:din_def}---legal due to Python's lexical scoping rules. In other words, the value of \pyi{Din} will come from the \emph{calling} context. In this case, \pyi{Din} on line~\ref{lne:uni_end} is replaced with the value defined on line~\ref{lne:din_def} due to \pyi{unify()} being accessed on line~\ref{lne:uni_call}. The code, though, results in the following (run-time) \pyi{NameError} on line~\ref{lne:uni_end}: \pyc{free variable 'Din' referenced before assignment in enclosing scope}~\cite{Dutordoir2020}. The problem is that, while it itself is not a \TFF*, \pyi{ndiagquad()} is \emph{called} by a \TFF* elsewhere---it will also be compiled into a (static) graph (cf.~\cref{sec:res:quant:others}). Thus,
dynamic language features like lexical scoping are not available in static contexts. As a result,
\pyi{unify()} is moved to line~\ref{lne:uni_new}, where \pyi{Din} is in its declaration scope. Although Python is a dynamic language, developers must be aware that certain code will be compiled to static graphs, leading to \cref{ap:dyn}, \cref{fig:aps}.

\subsubsection{API Misuse}\label{sec:res:qual:apm}

\begin{commit}{\addons, \href{https://github.com/tensorflow/addons/commit/8bab32264837494638da0f9a2b7652b282aac274}{\textinline{8bab3226}}, tensorflow_addons/image/dense_image_warp.py, Sean Morgan, seanmorgan91@gmail.com, Apr, 18, 2019, remove tf.func, apm}
    \begin{diffcode}
|\label{lne:remtff}|- @tf.function
  def interpolate_bilinear(grid, query_points, indexing="ij", name=None): # ...
|\label{lne:assert_begin}|  	tf.debugging.assert_equal(query_shape[2], 2, message=
|\label{lne:assert_end}|  		"Query points must be size 2 in dim 2.")
    \end{diffcode}
\end{commit}

On line~\ref{lne:remtff} in \cref{lst:apm}, \TFF\ is removed to fix a bug that is causing flaky tests~\cite{Morgan2019}. The problem is deemed to be that \TFF\ and the assert statement on lines~\ref{lne:assert_begin}--\ref{lne:assert_end} is incompatible. The developers express that ``removing the decorator is not ideal, but stability is more important than the [speedup] we [would] get with [it]''~\cite{Morgan2019a}. However, this code likely causes a race condition because of a missing control dependency following the assertion. To use the assertion within a \TFF*, a control dependency is required ``to block follow-up computation[s] until the check has executed'' as a result of the function being converted to a (static) graph~\cite{Google2021q}. This leads to \cref{bp:docs}, \cref{fig:bps}.

\subsubsection{TensorFlow Bugs}

\begin{commit}{\neuro, \href{https://github.com/MLH-Fellowship/neuro-art/commit/8a77761e1d52b7949283d154a29985d3a42347fe}{\textinline{8bab3226}}, backend/model/nst.py, David Knox, dakdave26@gmail.com, Nov, 11, 2020, Multiple request bugfix\ldots, tfb}
    \begin{diffcode}
|\label{lne:remtff2}|- @tf.function
  def train_step(image):
  	with tf.GradientTape() as tape:
  		outputs = extractor(image)
  		loss = style_content_loss(outputs)
  		loss += total_variation_weight * tf.image.total_variation(image)
  	grad = tape.gradient(loss, image)
|\label{lne:grads}|  	opt.apply_gradients([(grad, image)])
  	image.assign(clip_0_1(image))
    \end{diffcode}
\end{commit}

On line~\ref{lne:remtff2}, \cref{lst:tfb}, \TFF\ is once again removed. The problem is that---with \TFF---the application ``can only process one image before'' needing to restarted~\cite{Knox2020}, terminating with the message: \pyc{ValueError: tf.function-decorated function tried to create variables on non-first call}. Recall from \cref{sec:res:api} that shared variables inside a \TFF* must be singleton; a run-time exception ensues otherwise~\cite{Google2021b}.
However, it is not obvious from \cref{lst:tfb} where the variable creation occurs---there are no explicit \pyi{tf.Variable}s. The developer expresses that ``removing the \ldots\ decorator is a viable workaround but not [a] best practice,'' and that the root cause is an (unresolved) \citetitle{Abadi2016} bug~\cite{Stavarache2019}. In terms of \cref{lst:tfb}, the problematic line is~\ref{lne:grads}, as ``calling \pyi{apply_gradients()} on an optimizer for the first time will create its internal variables''~\cite{Passos2019}. In terms of the framework, it transpires to be related to
software layering,
as, ``sadly[,] there [is] currently no public API to just initialize the optimizer state but not [apply it].'' While several developers found workarounds for their particular situations, imperative DL code such as that in \cref{lst:tfb} have foregone any potential performance gains from using \TFF, leading to \cref{bp:bugs,ap:vars} in \cref{fig:bps,fig:aps}, respectively.

\subsubsection{Debuggability}\label{sec:res:qual:dbg}

\begin{commit}{\addons, \href{https://github.com/tensorflow/addons/commit/16ee6c599daf52a9dc8024a4615f96f1e2bbce9e}{\textinline{16ee6c59}}, tensorflow_addons/layers/wrappers.py, Qianli Scott Zhu, scottzhu@google.com, Apr, 25, 2019, tf.func for debug, dbg}
\begin{diffcode}
|\label{lne:remtff3}|- @tf.function
  def call(self, inputs):
  	"""Call `Layer`"""
- 	if not self.initialized:
- 		self._data_dep_init(inputs)
+ 	if not self._initialized:
+ 		self._initialize_weights(inputs)
  	self._compute_weights() # Recompute weights for each forward pass ...
\end{diffcode}
\end{commit}

To improve debuggability, \TFF\ is removed on line~\ref{lne:remtff3}, \cref{lst:dbg} (cf.~\cref{sec:res:quant:dbg}). However, in the latest file version,
\TFF\ has not been replaced. Thus, the developer may have inadvertently sacrificed permanent performance gains for temporary debuggability, leading to \cref{bp:dbg}, \cref{fig:bps}.

\section{Discussion}\label{sec:discuss}

We summarize and comment on our main findings while connecting them to other research.
To help solve the problems, we put forth preliminary recommendations for practitioners, tool developers, and researchers. Though our hope is that the findings will shed light on future tool challenges and that the aforementioned descriptions and real-world examples will provide sufficient, generalizable, and actionable contexts, we nonetheless outline potential solutions.

\begin{figure}[t]
    \small
    \begin{recommendations}
	\item
	    More tool-support
	    for assisting with using \mintinline[fontsize=auto]{python}{tf.function} may help produce reliable yet performant imperative DL code.\label{suggest:prf}
	\item Modernize and reformulate existing tensor shape mismatch detectors
	    for imperative DL code and \mintinline[fontsize=auto]{python}{tf.function(...)} input shapes.\label{suggest:ins}
	\item More formal specification in a design-by-contract (DbC) style may be helpful for new tool-support aimed to alleviate API misuse.\label{suggest:dbc}
	\item Testing (dynamic analysis) focused on hybridized (imperative) DL code that runs under \emph{multiple} execution modes may
	    localize bugs.\label{suggest:test}
    \end{recommendations}
    \caption{Preliminary hybridization recommendations.}\label{fig:suggest}
\end{figure}

\textit{Performance. }\label{sec:discuss:dup}
It is not surprising that performance is our largest category (\cref{fnd:prf}) since hybridization
is centrally related to performance enhancement. The volume of performance problems is a testament to the struggles developers have in writing performant, imperative DL code. However, \PerfFixedOtherPercentage\ of performance problems (\cref{fnd:prffixes}) were due to \emph{existing} \TFF*\ usages, suggesting that developers also struggle with using hybridization \emph{effectively} to achieve the performance they desire. A feasible explanation is that developers must \emph{manually} decide:
\begin{enumerate*}[(i)]
    \item where and when to use \TFF*,
    \item the arguments to supply \TFF*\ for their code to perform optimally, and
    \item which code is amenable to (efficient) graph conversion and which is not,
\end{enumerate*} all of which can be error-prone.

The overarching goal is reliable \emph{and} performant DL code; reliability stems from being able to write DL code in a less error-prone imperative-style, while performance is achieved in migrating that code to graph execution at run-time. \citetitle{Moldovan2019}, as well as other hybridization technologies, attempt to achieve this goal by automating the migration process as much as possible---frequently requiring contextual information from developers as to their intentions and imposing limitations of where the technology can be used. The end result is a trade-off---one of many typically made by DL frameworks~\cite{Islam2019a}. As discussed in \cref{sec:intro}, others~\cite{Jeong2019} attempt to automate the entire migration process---not requiring any contextual metadata---but impose new trade-offs, such as necessitating custom Python interpreters that may not be practical for industrial applications and support only specific Python constructs.

As \citetitle{Moldovan2019} and other hybridization technologies are pervasively used, as well as being integrated into official distributions of popular DL frameworks, our suggestion is to retain (and continually improve upon) hybridization platforms, while simultaneously posing this problem as one of (API) \emph{usability}. Our perspective is that what is needed it tool-support that will guide developers in using this technology correctly given a particular context, as well as automated refactoring and other source code transformation tools that can detect and repair hybridization problems. Such techniques would alleviate hybridization issues well-before they are seen beyond (production) deployment or after long training sessions, leading to \cref{suggest:prf}, \cref{fig:suggest}.

\citet{Zhang2018,Cao2021} also study
performance of DL code and found that
a modest portion of \emph{non-imperative} \citetitle{Abadi2016} program bugs involved performance problems. However, these problems were caused by confusion with the underlying computation model, which essentially requires developers to build graphs \emph{manually}. In our case, graphs are built \emph{automatically}. Since imperative DL code runs eagerly by default, it is understandable that our study
would uncover more performance problems. In fact, \citet{Tambon2021} also observe performance degradation of imperative DL code.
\citet{Wan2017} found that performance bugs took the longest average time to fix in blockchain systems.

Per \cref{fnd:perfargfix}, developer-supplied arguments to \TFF*\pyi{()} played a major role in performance problems, comprising \PerfFixedByArgsPercentage\ of performance fixes. Furthermore, per \cref{fnd:shapes}, a significant percentage (\InputShapePercentageOfPerformanceProblems) of performance problems involved parameters representing input shape specification---one of the most frequently used \TFF*\ parameters (q.v.~\cref{sec:res:quant:ins}). Input shape problems are a central focus of related work~\cite{Lagouvardos2020,Zhang2018,Islam2019} on DL programs; related studies~\cite{Zhang2018,Islam2019,Islam2020a,Humbatova2020} also found shape problems.
A feasible explanation is that developers are challenged to determine tensor shapes from all possible call sites statically. We again advocate for more tool-support in this area, e.g., an adaptation of \citet{Lagouvardos2020} for imperative DL programs focused on hybridization parameters, leading to \cref{suggest:ins} in \cref{fig:suggest}.

\textit{API Misuse. }
Per \cref{fnd:apm}, using the \TFF*\ API inconsistently with its documentation was a major theme. Feasible explanations include:
\begin{enumerate*}[(i)]
    \item DL APIs---along with their documentation~\cite{Hashemi2020}---are particularly vast and complex~\cite{Islam2019a},\label{itm:complex}
    \item often, documentation consumers (developers) are not software experts~\cite{Hashemi2020},\label{itm:exp}
    \item although developers are writing imperative DL code, there exist situations where they must nevertheless be cognizant of hybridization limitations, and\label{itm:limits}
    \item error messages may not be helpful.\label{itm:err}
\end{enumerate*}
Due to \cref{itm:complex}, learning how to use DL APIs effectively necessitates a steep learning curve, especially considering that hybridization is
relatively new.
As ML systems have a quick time-to-market~\cite{Sculley2015}, developers may be not have the luxury of time to thoroughly understand the documentation. This is especially evident in \cref{fnd:apmcause}, with \APIConfusionPercentage\ of misuses caused by API confusion. \Cref{itm:exp} has been recognized by other ML/DL software studies (e.g.,~\cite{Tang2021}). We conjecture that \cref{itm:limits} can also be alleviated with more tool-support, however, such tool-support in this context may require (e.g., design-by-contract) formalization of DL API specifications (e.g.,
modeling operation limitations in particular contexts), leading to \cref{suggest:dbc}, \cref{fig:suggest}. A potential downside to \cref{suggest:dbc}
is the rapid change of ML APIs~\cite{Dilhara2021}. For \cref{itm:err},
developers often expressed frustration with error messages, e.g.,  ``the main complexity in [\citetitle{Abadi2016}] 2 is in \TFF[;] \ldots\ error messages should be as clear as possible, especially for common problems''~\cite{Geron2019a}.

\citet{Zhang2019} likewise observed broader API misuse in DL systems. \citet{Nadi2016} also found API misuse despite ample documentation in the context of cryptography---developers prefer higher-level documentation. Current hybridization documentation tends to focus on lower-level details---future research may explore whether a similar concept will work for DL APIs. Furthermore, our findings coincide with \citet{Jin2012} that many performance bugs are due performance implication misunderstandings of certain functions.

\textit{Incompatibility.}
Execution incompatibility of particular Python constructs
was also a major theme (q.v.~\cref{fnd:inc}). %
\citet{Zhang2019} found a similar problem in DL systems w.r.t.~CPU/GPU compatibility.
We again advocate for more automation
to circumvent such problems. To use hybridization effectively, developers must understand which constructs are amenable to \emph{both} eager and graph execution and make appropriate considerations. Tool-support,
e.g., IDE
recommendations, may be helpful here.
To alleviate run-time errors and unexpected results, we also advocate for more testing (dynamic analysis) of (imperative) DL code
that runs the same code under \emph{multiple} execution modes. Testing of DL systems is an emerging yet promising area, and testing focusing on (imperative) DL code hybridization may help to shed light on:
\begin{enumerate*}[(i)]
    \item where developers struggle to write performant yet reliable (imperative) DL code and
    \item potential areas of where hybridization technologies can be improved.
\end{enumerate*}
This leads to \cref{suggest:test}, \cref{fig:suggest}.

\textit{Commits vs.~\GH\ Issues. }\label{sec:discuss:vs}
Performance bugs appeared more in commits than \GH\ issues. The reason may be that enhancing performance typically requires a code change, which can be benchmarked. Contrarily, ``incompatibility'' is more difficult to quantify, often resulting in unexpected behavior or run-time errors (q.v.~\cref{fig:inc_symptoms}). Therefore, developers may be more likely to seek external assistance.
Developers commonly file \GH\ issues against \citetitle{Abadi2016}; \SI{93.75}{\percent} of TFB issues are against the \TF\ subject. That all UKN and TST bugs appeared in commits may be due to \GH\ issues
being easier to categorize than changesets and DL testing remains an emerging area, respectively.

\section{Threats to Validity}\label{sec:threats}

Subjects may not be representative of DL systems. To mitigate this, subjects encompass diverse domains and sizes, have been used in previous studies, and are from a data science-specific dataset
(q.v.~\cref{sec:subj}). Various GitHub metrics and DL-related keywords were used
in choosing subjects.
Also, hybridization is relatively new; we expect a larger selection of subjects as it grows in popularity.

Our study involved many hours of manual validation to understand and categorize bugs. To mitigate bias, we investigated referenced resources and comments made by developers to help more fully understand the challenges faced. The NLP
of \citetitle{Casalnuovo2017} may have missed bug fix changesets.
Nevertheless, using it, we were still able to find \NumBugs\ bugs (\NumChallenges\ overall) that contributed to a rich bug categorization,
best practices, and anti-patterns. Furthermore, \citetitle{Casalnuovo2017} has been used previously in other studies (q.v.~\cref{sec:meth:mine}). %

Hybridization
in comparable DL frameworks
may have yielded different challenges.
Nevertheless, focusing on
\citetitle{Abadi2016}
enables us to more thoroughly understand the intricacies involved in using
hybridization effectively. Moreover, \citetitle{Abadi2016} is a widely-studied and popular (industrial) DL framework (q.v.~\cref{sec:intro}).

\section{Related Work}\label{sec:related}

\citet{Cao2021} characterizing performance bugs in DL systems. During their analysis of general performance bugs, they also find that developers often struggle with knowing where to add \TFF\ and how to implement decorated functions for optimal performance. Beyond performance bugs, our study includes a rich, hierarchical taxonomy of varying hybridization bug types, including input shape mismatches, API misuse, and construct incompatibility, whose results include run-time errors, unexpected behavior, and deadlock. \citet{Tambon2021} examine (silent) behavioral bugs \emph{within} DL frameworks and their impact on client code. Their work is reminiscent of our TFB problem category (q.v.~\cref{sec:res:tfb}) and also note that performance degradation may lead to significant problems at run-time. While they do not explicitly mention hybridization performance bugs, some of their performance bugs in imperative DL code may be alleviate by using \TFF. \citet{Zhang2021} study API change trends in \citetitle{Abadi2016} and for which reasons; our focus is on \emph{client} code modifications involving hybridization. \citet{Baker2022} extract \num{11} common \citetitle{Abadi2016} API misuse patterns. Only one of the patterns (and corresponding fix suggestion) involves (a specific use case of) \TFF*. In contrast, our study goes beyond API misuse and entails \TopLevelCategories\ top-level problem categories---\OverallCategories\ overall---encompassing hybridization challenges. %

\citet{Zhang2019} present a large-scale empirical study of general DL questions on Stack Overflow. Particularly, their ``CPU/GPU incompatibility'' problem category resembles our execution mode incompatibility category. Concerning hybridization, whether the migrated graph
executes on a GPU is typically decided by the underlying DL framework;
our focus is on conversion itself. \citet{Islam2019,Zhang2018} study general DL bug characteristics and present anti-patterns to avoid bugs. \citet{Islam2020a} study patterns in which such bugs are fixed. \citet{Chen2021} explore faults in deploying DL models to mobile applications. \citet{Nikanjam2021} catalog various design smells in DL systems and recommend suitable refactorings. \citet{Jebnoun2020} correlate code smells with bugs in DL code. \citet{Liu2020} characterize technical debt in DL frameworks, while \citet{Humbatova2020} taxonomize (functional) faults in DL systems. \citet{Arpteg2018} categorize (general) SE challenges in DL systems into three areas---development, production, and organizational.
\citet{Liu2021} study failed \citetitle{Abadi2016} industrial jobs and propose a constraint-based approach for detecting shape-related errors. \citet{Amershi2019} conduct a study at Microsoft, observing software teams as they developed AI applications.
\citet{Lwakatare2019} also classify SE challenges for ML systems at six different companies, focusing mainly on deployment issues. \citet{Thung2012} examine bugs in three general ML systems, finding that nonfunctional bugs, of which performance problems may be categorized, require the most involved fixes. \citet{Dilhara2021} study ML library evolution and its resulting client-code modifications. And, \citet{Dilhara2022,Tang2021} analyze repetitive code changes and refactorings made in ML systems, respectively. While valuable, these studies do not deal with challenges faced in migrating imperative DL code to graph execution.

Several studies involve performance in other contexts. \citet{Han2016} study
configurability and performance.
Future work entails correlating their findings with \TFF*\ arguments.
\citet{Jin2011} study performance slowdowns caused by system side inefficiencies.
\citet{Bagherzadeh2020} investigate performance in Actor-based systems.
Others
study language features. \citet{Parnin2013} study Java generics adoption. \citet{Dyer2014}
study language feature evolution. \citet{Khatchadourian2018}
empirically assess
default methods.
There are many general empirical studies.
\citet{Makhshari2021} taxonomize development challenges of IoT systems.
\citet{Bagherzadeh2019} investigate common questions asked by big data developers, and \citet{Khatchadourian2020} examine the use and misuse of Java streams.
\citet{Engler2001,Tian2017} study errors in systems code.

\section{Conclusion \& Future Work}\label{sec:conc}

This study advances knowledge of the development challenges involved in migrating imperative DL code to graph execution via hybridization.
A hierarchical taxonomy of common hybridization challenges was formulated and preliminary recommendations, best practices, and anti-patterns were proposed. In the future, we will explore analyzing alternative developer resources, e.g., Stack Overflow, and integrating our results into automated bug finders and refactoring detection approaches~\cite{Tsantalis2020,Atwi2021}.

\printbibliography%

\end{document}